\documentstyle[12pt]{article}
\def \unit {{  1 \kern -0.27em {\rm I}}\,}
\def \MSbar {\vbox{\hrule\kern 1pt\hbox{\rm MS}}}

\def\lesssim{\ \hbox{\raise 2pt \hbox{$<$} \kern -13pt
                     \lower 3pt \hbox{$\sim$}}\ }
\def\greatersim{\ \hbox{\raise 2pt \hbox{$>$} \kern -13pt
                     \lower 3pt \hbox{$\sim$}}\ }

\input epsf.tex
\def\desepsf(#1 width #2){\epsfxsize=#2 \epsfbox{#1}}
\begin{document}
\begin{titlepage}
\renewcommand{\thefootnote}{\fnsymbol{footnote}}

\vspace*{5mm}
%\par \vskip 10mm

\begin{center}
{\Large \bf Cancellation of Sudakov Logarithms in Radiative 
 Decays of Quarkonia} \\
%\end{center}
\vspace*{1cm}

        \par \vskip 5mm \noindent
        {\bf F.\ Hautmann}\\
        \par \vskip 3mm \noindent              
        Department of Physics, Pennsylvania State University, 
        University Park PA 16802\\

\par \vskip 1cm

\end{center}
%\par \vskip 2mm
\vspace*{1cm}

\begin{center} {\large \bf Abstract} \end{center}
\begin{quote}
We study infrared QCD effects in  radiative quarkonia decays. 
 We examine the  endpoint region $z \to 1$ of the photon spectrum. 
We point out a cancellation mechanism for the corrections 
 in $\alpha_s^n \ln^{m} (1-z)$, $m \leq 2 n$, 
 in the short-distance coefficient  for  the 
color-singlet Fock  state  in the quarkonium. The cancellation is due 
to the coherence of the  color radiation, and applies even though 
 logarithmic contributions are present in 
the jet distributions associated with the decay.       
  We comment on the implications of our results 
for the modeling of hadronization  in the endpoint region 
and  for  the role of color-octet  states  in the quarkonium. 
\end{quote}
\vspace*{3.8 cm}

\begin{flushleft}
     PSU-TH/236\\
     February 2001\\
\end{flushleft}
\end{titlepage}

\renewcommand{\thefootnote}{\fnsymbol{footnote}}

\noindent  {\large \bf 1. Introduction} 
\vskip .2 true cm

 Radiative decays of 
heavy quark-antiquark bound states (quarkonia) 
have been investigated in the framework of  perturbative QCD    
and  
have been used 
to measure   
the QCD coupling at energies of the order of the heavy quark 
mass $m$~\cite{kobel,montfield,cleo,field01}.  
The use of perturbative QCD 
in this context  is based on the fact that, as long 
as the velocity $v$ of the heavy quark  is small, these decay processes 
involve two widely separated distance scales: the  
scale $1/ (m v^2)$ over which the quark and antiquark bind into the  
quarkonium and the 
scale $1/m$ over which the quark-antiquark pair decays. 
By expanding about the nonrelativistic limit $v \to 0$, one may treat 
the process as  the  product of a long-distance, 
nonperturbative factor, containing all of 
the bound-state dynamics, and a short-distance factor, 
 describing the annihilation of the  
heavy quark pair and  computable as a 
power series expansion in $\alpha_s$. 

However, near the exclusive boundary of the phase space, 
 where the photon's energy 
 approaches its kinematic limit, 
both the expansion to fixed 
order in $v$ and the expansion to fixed order in $\alpha_s$ 
become inadequate  to represent correctly the physics of the decay. 
On one hand,  classes of relativistic corrections  
 that  by power counting are higher order 
in $v$   become enhanced by 
powers of $1 / \alpha_s$~\cite{rothwise,malpet,wolf}. 
On the other hand,  
potentially large terms in $\ln (1-z)$ (with  
 $1-z$ being 
 the parameter that measures the distance from the endpoint)  
appear in the coefficients   of  
the expansion in $\alpha_s$ to all orders~\cite{Photia}.  In both cases, 
reliable results can only be obtained after  resummation of the enhanced 
contributions. 

Infrared QCD effects are responsible for these behaviors.    
Soft color interactions account
for   the heavy quark-antiquark pair 
turning into a color-octet state, whose 
contribution, although subleading in $v$, becomes important near the 
endpoint. Sudakov 
logarithms in  $1-z$ arise as remnants of the imperfect cancellation 
of the infrared divergences  in the exclusive  region, where 
particle production  is suppressed. 
The need for resummations noted above is  
a symptom of the sensitivity  
of the process 
to contributions from infrared scales. 
A corresponding effect is seen in  calculations that 
incorporate  models for  
the hadronization of partons~\cite{field01,Field,consfie}, where      
nonperturbative contributions are found to be  essential to 
describe the endpoint region of the photon energy spectrum in 
$J/\psi$~\cite{mark2} and $\Upsilon$~\cite{cleo,earlierups} decays.  

In this paper we present a treatment of the infrared QCD radiation 
near the kinematic boundary. We describe contributions from 
soft and collinear gluons with logarithmic accuracy, and 
study the 
 color correlations between 
 the jets 
that accompany the photon. 
 This 
treatment  takes account of 
 the coherence properties of 
 gluon emission.  
   We determine the regions in the 
final-state phase space 
for   the quarkonium decay 
in which destructive interference between parton emitters 
suppresses color radiation. 

We  use these results  to examine 
potentially large Sudakov logarithms in the photon spectrum. 
We find 
(see Sec.~4) that 
Sudakov logarithms cancel order by order in $\alpha_s$ in the short 
distance coefficient for the color-singlet Fock state in the 
quarkonium. 
The cancellation mechanism 
applies to the leading logarithms as well as 
to any subleading logarithms.  
As a result, no Sudakov suppression factor arises 
near the endpoint. 

The picture underlying 
this behavior  can be understood 
in simple terms based on the 
branching formulas for the decay  discussed  
 in Sec.~2. 
In the boundary kinematics 
the photon recoils against two almost-collinear gluon jets. 
 The cancellation of Sudakov corrections  reflects the fact that 
color is neutralized already at the level of this two-jet configuration.  
  This situation may be   
  contrasted with the situation one encounters in the 
  decay of an electroweak gauge  boson into jets. 
Here  Sudakov corrections arise precisely from the presence of color 
charges in two-jet configurations. 

This picture also indicates that no  cancellation should 
occur  for the color-octet Fock state in the quarkonium. In this case, 
we expect the usual Sudakov suppression to take place near the 
endpoint of the photon  spectrum. Then one of the consequences 
of the results of this paper concerns the ratio of the  
 color-octet   to the  color-singlet contributions. 
 This ratio  will be smaller than expected from the 
power counting  in $v$ and $\alpha_s$ 
obtained by truncating perturbation theory
to  fixed  order~\cite{rothwise,mannwolf,grekap}, owing to the 
different high order  behavior of the perturbation 
series  in the two cases.   

A preliminary discussion of the results presented in this paper 
appeared in Ref.~\cite{egmond}.

The paper is organized as follows. In Sec.~2 we  treat the emission of 
  soft and collinear gluons  associated with the quarkonium decay.
This leads us to 
 a general branching formula.  
In Sec.~3 we evaluate 
the phase space in  this formula. 
In Sec.~4 we  compute the photon spectrum and 
observe the cancellation of the Sudakov logarithms. In Sec.~5 
 we discuss the implications of our results and give conclusions.

\vskip 1.5 true cm 

\noindent  {\large \bf   2. Infrared  radiation 
near the phase space boundary } 
\vskip .2 true cm

In this section  we describe  the  QCD radiation  
that accompanies the  decay of 
the quarkonium  near the endpoint of the photon 
 spectrum. The key element  in our treatment is  
 the coherent emission of gluons and 
 the angular ordering of this emission. 
 We arrive at an expression for the photon spectrum that 
 is valid to leading and next-to-leading accuracy in the soft 
and/or collinear logarithms.

We consider the  decay of a 
quarkonium state $H$ of mass $M$ into a photon 
plus anything:  
\begin{equation}
\label{raddec}
H  \to \gamma + X  
\hspace*{0.5 cm} . 
\end{equation} 
We concentrate on the case of 
the leading Fock state 
for a $^{3} S_1$ quarkonium state, that is,  a heavy quark-antiquark 
pair in a color singlet state.  

This decay     
receives contributions  at the leading 
order of perturbation theory   
both when the photon is directly coupled to the heavy quarks  
(direct term) 
and 
  when the 
 photon is produced by collinear emission from 
 light quarks  (fragmentation term)~\cite{mont}. 
In the endpoint region  
the fragmentation contribution 
 is suppressed relative to the direct term. The discussion of this 
 paper will thus 
 focus on  the direct term.

\vskip 0.7 true cm 

\noindent  {\it 2.1 Notations and  tree-level decay  } 
\vskip .2 true cm

The direct contribution  at the leading order 
 is given  by the process  
\begin{equation}
\label{borndir}
H (P) \to \gamma(k) + g(k_1) + g(k_2) 
\hspace*{0.5 cm}  
\end{equation}
evaluated at the tree level. 
We  introduce dimensionless energy variables for the particles in the 
final state,  as follows 
\begin{equation}
\label{kindef}
z = {{ 2 P \cdot k} \over {M^2} } \hspace*{0.1 cm} , 
\hspace*{0.3 cm} x_i = {{ 2 P \cdot k_i} \over {M^2} } \hspace*{0.1 cm} , 
\hspace*{0.3 cm} i = 1,2 
\hspace*{0.6 cm} . 
\end{equation}
In the quarkonium rest frame $z$ and 
$x_i$ are  respectively the photon and gluon energies 
scaled by the heavy quark mass $m$ ($m \simeq M/2$ 
in the lowest nonrelativistic approximation).

 The  normalized photon spectrum   from 
  the process (\ref{borndir})   is given by~\cite{Brod}  
\begin{equation}
\label{borng}
{ 1 \over \Gamma_0} \ {{d \Gamma_0} \over { d z} } = 
{1 \over {\pi^2 -9}} \ \int_0^1 d x_1 
  \int_0^1 d x_2   \ \rho (x_1 , x_2, z) \ \delta(z - 2 + x_1 + x_2)
\  \Theta(x_1+x_2 -1) 
\hspace*{0.1 cm} , 
\end{equation}
with 
\begin{equation}
\label{tree}
\rho (x_1 , x_2, z) = 
  {{(1-x_1)^2} \over { z^2 x_2^2}} + 
{{(1-x_2)^2} \over { z^2 x_1^2}} + {{(1-z)^2} \over { x_1^2 x_2^2}} 
\hspace*{0.4 cm} .  
\end{equation}
Over almost the entire range   in $z$ 
 the result (\ref{borng}) is well 
 approximated by a spectrum rising linearly with $z$. At the 
 endpoint, however,  
 the first derivative of the spectrum (\ref{borng}) is singular. 
 We will come back to this in Sec.~4.

\vskip 0.7 true cm

\noindent  {\it  2.2  Behavior at $z \to 1$
} 
\vskip .2 true cm

Beyond the tree level, 
  corrections in  $\ln (1-z)$ to the   spectrum 
 become possible~\cite{Photia}. 
 These come  from   the infrared region: near the endpoint  
 the emission of gluons is suppressed and terms in $\ln (1-z)$ 
result from  the imperfect 
cancellation of the infrared divergences between real and 
 virtual graphs. The contribution from 
 the  soft and/or collinear gluons associated with the decay 
is at most double-logarithmic for each power of $\alpha_s$. 
 By power counting 
 the  leading behavior of  the photon spectrum  
as $z \to 1$ is   of the type~\cite{Photia}  
\begin{equation}
\label{hologs}
{1 \over \Gamma} \ 
{{d \Gamma} \over { d z} } \sim 
{\mbox{const.}} \ + \sum_{k=1}^{\infty} 
c_k \ \alpha_s^k \ \ln^{2 k} (1-z) \;\; ,  \;\;\;\;\; z \to 1 
\hspace*{0.4 cm} . 
\end{equation}
 
 In this paper we will see a cancellation mechanism for 
 these corrections that works order by order  
in perturbation theory. A qualitative picture of the cancellation that    
we will discuss in the next sections can be given at one loop as follows. 
 Consider the  emission of 
a soft gluon with momentum $q$ 
from the tree-level graphs for the decay (\ref{borndir}).   
In the 
leading infrared 
approximation,  
the amplitude ${\cal A}$ for this process 
factorizes  (see for instance \cite{coh}) into 
 a nonabelian, classical current 
 ${\bf j}(q)$,  describing the soft-gluon emission,  
 times the tree level amplitude ${\cal A}_0$:   
\begin{equation}
\label{softampl}
{\cal A}^b (P, k_1, k_2; q) \simeq 
g_s \varepsilon^\mu (q) {\bf j}_\mu^b (q) {\cal A}_0 (P, k_1, k_2)
 \hspace*{0.4 cm} , 
\end{equation}
where $\varepsilon$ is the gluon polarization vector, $b$ is the 
gluon color index,  
and 
\begin{equation}
\label{eikcurr}
{\bf j}^\mu  (q) = \sum_i {\bf T_i} {k^\mu_i \over {k_i \cdot q}} 
\hspace*{0.4 cm} .  
\end{equation}
Here  the sum runs over the 
final-state colored particles $k_1$ and $k_2$, and 
${\bf T_i}$ are the  color matrices associated with the coupling of 
the soft gluon to 
each of these particles. 
 There is no contribution from the coupling to quarks: 
the coupling to the virtual quark lines 
is subleading in the infrared limit because these  lines 
are 
off shell by order $M$; 
 the  
coupling to the  
 quark and antiquark in the quarkonium does not contribute either, 
although these   quarks  are nearly on shell,  because in the lowest 
nonrelativistic approximation they are almost collinear and their 
total color charge is zero.

To the  approximation 
at which we are working,   
the phase space also  factorizes~\cite{coh}, 
so that one gets a factorized 
answer for the  spectrum, schematically of the form
\begin{equation}
\label{oneloop}
{{ d \Gamma} \over {  d z }} \simeq \alpha_s \int  
{\bf j}^2 \, 
{{d \Gamma_0} \over { d z} } 
\, d \Phi^\prime \hspace*{0.4 cm} .  
\end{equation}
Here $d \Phi^\prime$ is the Lorentz-invariant 
   phase space for the soft gluon emission.  
The standard power counting 
in terms of the soft gluon energy $\omega$ gives  
\begin{equation}
\label{phij}
d \Phi^\prime \sim \omega \, d \omega \, d \Omega \hspace*{0.2 cm} , 
\hspace*{0.4 cm} {\bf j}^2 \sim { 1 \over {\omega^2}} \, {{ k_1 \cdot k_2 }
\over { f (z ; {\mbox {angles}} )}} \hspace*{0.4 cm} ,  
\end{equation}
where $ d \Omega$ is the angular phase space, and $f$ is  a 
function of 
$z$ and 
the angles as $\omega \to 0$. 
Up to the first order in $\omega$ 
the  correlation of the gluon momenta is 
\begin{equation}
\label{k12}
 k_1 \cdot k_2 \sim M^2 \left[ (1 - z) + {\omega \over M} \, 
g (z ; {\mbox {angles}}) \right] \hspace*{0.4 cm} .
\end{equation}
That is, as $z \to 1$  
the photon recoils against two almost-collinear hard gluons. 
In this configuration, 
the logarithmic integration 
 $ d \omega  /  \omega $ in Eq.~(\ref{oneloop}) 
is canceled. 
The next few sections are devoted to showing that this is 
indeed the mechanism that dominates the radiation 
accompanying the decay, and  
 that 
 it extends to all orders in $\alpha_s$. 

The back-to-back kinematics  suggests an 
analogy with the two-jet region in the  annihilation of 
$e^+ e^-$ into 
electroweak gauge bosons.  
Since the  two jets are  now 
 in a singlet, though, the pattern 
of color correlations  will be  different. Jet event shapes 
in $e^+ e^-$  annihilation have been studied extensively by 
using methods based on 
 branching graphs. 
We will see in the 
remainder  of this section that 
the photon spectrum in decays of 
quarkonia can be treated in a similar manner. 
The subtlest point will be the 
treatment of the branching  phase space. This  will be addressed 
in Sec.~3.

\vskip 0.7 true cm

\noindent  {\it  2.3  Branching graphs} 
\vskip .2 true cm

Methods to calculate multi-parton matrix elements in the leading soft 
and collinear orders have been known for a long time. See 
Ref.~\cite{coh} for a review. The essential observation in these 
methods is that one can replace the calculation of higher-loop Feynman 
graphs by the calculation of angular-ordered ``branching'' graphs, 
that is, tree level graphs in which the angular phase space is subject 
to certain ordering constraints at each branching. The method has also 
been extended to include next-to-leading logarithms: see for instance 
Ref.~\cite{cttw}. In this and 
the next subsection we apply this approach to the 
quarkonium decay. 

We write the   energy spectrum  $d \Gamma / dz$ 
in the inclusive decay 
(\ref{raddec}) in terms of the exclusive 
decay widths $d \Gamma_n^{(\rm{excl})}$ for producing 
$\gamma$ $+$ $n$ final state partons $q_1, \dots, q_n$: 
\begin{equation} 
\label{excl}
{ 1 \over \Gamma} \ {{d \Gamma} \over { d z} } = 
\sum_{n=2}^{\infty} { 1 \over \Gamma} \int d \Gamma_n^{(\rm{excl})} 
(q_1, \dots, q_n) \delta \left( z - 2 + 
\sum_i {{ 2 P \cdot q_i} \over {M^2} } \right)
 \hspace*{0.2 cm} .      
\end{equation}
Here the $\delta$ function expresses the photon energy fraction $z$ in 
terms of the quarkonium and parton momenta. The 
 exclusive decay widths are generated by the branching 
process~\cite{coh,cttw,fad83}. 

The first step of the branching  corresponds to the lowest order 
final state,  Eq.~(\ref{borndir}), in which  
the photon is accompanied by two gluons with invariant 
mass $(k_1+k_2)^2$. This invariant mass provides the basic mass scale at  
 the hard end of the branching. 
The subsequent steps correspond to 
consecutive  parton splittings (Fig.~1a). 
There are well-defined probability weights 
associated with  each splitting vertex, and  form factors 
associated with each line connecting two vertices~\cite{coh,cttw}. 
The basic point about the branching  
process is that the phase space for  the splittings is restricted to the 
angular ordered region in which the branching angles decrease as 
we go from the parent parton toward the final state. If we denote, 
as in Fig.~1a, by 
$y_i$ the energy transfer at the vertex $i$, and by 
$r_{\perp i}$ the transverse momentum flowing between vertices 
$i$ and $i+1$, 
this region is defined (for small $y_i$) by      
\begin{equation} 
\label{angord}
(k_1+k_2)^2 \greatersim  {{ r_{\perp 1}^2} \over y_1^2} \greatersim 
{{ r_{\perp 2}^2} \over {y_1^2 \ y_2^2} } \greatersim \dots 
 \hspace*{0.2 cm} .      
\end{equation}
The reason for the angular ordering is a coherence effect: 
 outside the ordered region 
parton emitters interfere destructively~\cite{fad83,muecoh,dokfad}. 
In this approach, the exclusive decay width 
$d \Gamma_n^{(\rm{excl})}$ is obtained by taking  all 
angular-ordered branching graphs with $n$ final state partons, and summing 
the corresponding probability weights.

Following \cite{cttw}, we  integrate the exclusive decay widths 
over the phase space of the final state partons keeping 
$k_1$ and $k_2$ fixed. 
We rewrite the $n$-parton phase space as (Fig.~1b)
\begin{eqnarray}
\label{insert}
  \left[ \prod_{j=1}^{n} 
   {{ d^4 q_j} \over {(2 \pi)^3}} \delta_+ (q_j^2)  \right] 
  &=& 
  d^4 k_1 \delta^4 \left( k_1 - \sum_{j=1}^{l} q_j  \right)  
 \left[ \prod_{j=1}^{l} 
  {{ d^4 q_j} \over {(2 \pi)^3}} \delta_+ (q_j^2)  \right]  
\nonumber\\     &\times&
  d^4 k_2 \delta^4 \left( k_2 - \sum_{j=l+1}^{n} q_j  \right)  
 \left[ \prod_{j=l+1}^{n} 
  {{ d^4 q_j} \over {(2 \pi)^3}} \delta_+ (q_j^2)  \right]  
   \hspace*{0.4 cm} .  
\end{eqnarray} 
The integration of the branching matrix element over the 
final state phase space at fixed $k_i^2$ ($i = 1 , 2$) produces the 
jet mass distribution $J_g (p^2 , k_i^2)$~\cite{cttw}. This is   
 defined as 
the probability to generate 
a jet with invariant mass $k_i^2$ from a parent gluon produced at the 
mass scale $p^2$. We have  
\begin{eqnarray}
\label{defstruc}
 J_g \left( (k_1+k_2)^2 , k_1^2 \right) &=&  
 \left[ \prod_{j=1}^{l}  
 {{ d^4 q_j} \over {(2 \pi)^3}} \delta_+ (q_j^2)  \right] 
(2 \pi)^4 \delta^4 \left( k_1-\sum_{j=1}^{l} q_j \right)  
\nonumber\\
&\times& 
( {\mbox{branching}} \hspace*{0.2 cm} {\mbox{matrix}} 
 \hspace*{0.2 cm} {\mbox{element}} )
 \hspace*{0.4 cm} ,    
\end{eqnarray}
and an analogous expression for $J_g \left( 
 (k_1+k_2)^2 , k_2^2 \right)$.
 Note that the jet mass 
distributions depend on the mass scale  $(k_1+k_2)^2$. 
As implied by the coherence relation (\ref{angord}), 
 this is the mass scale  
at which the initial gluons in the branching are produced. 

\begin{figure}
\centerline{\desepsf(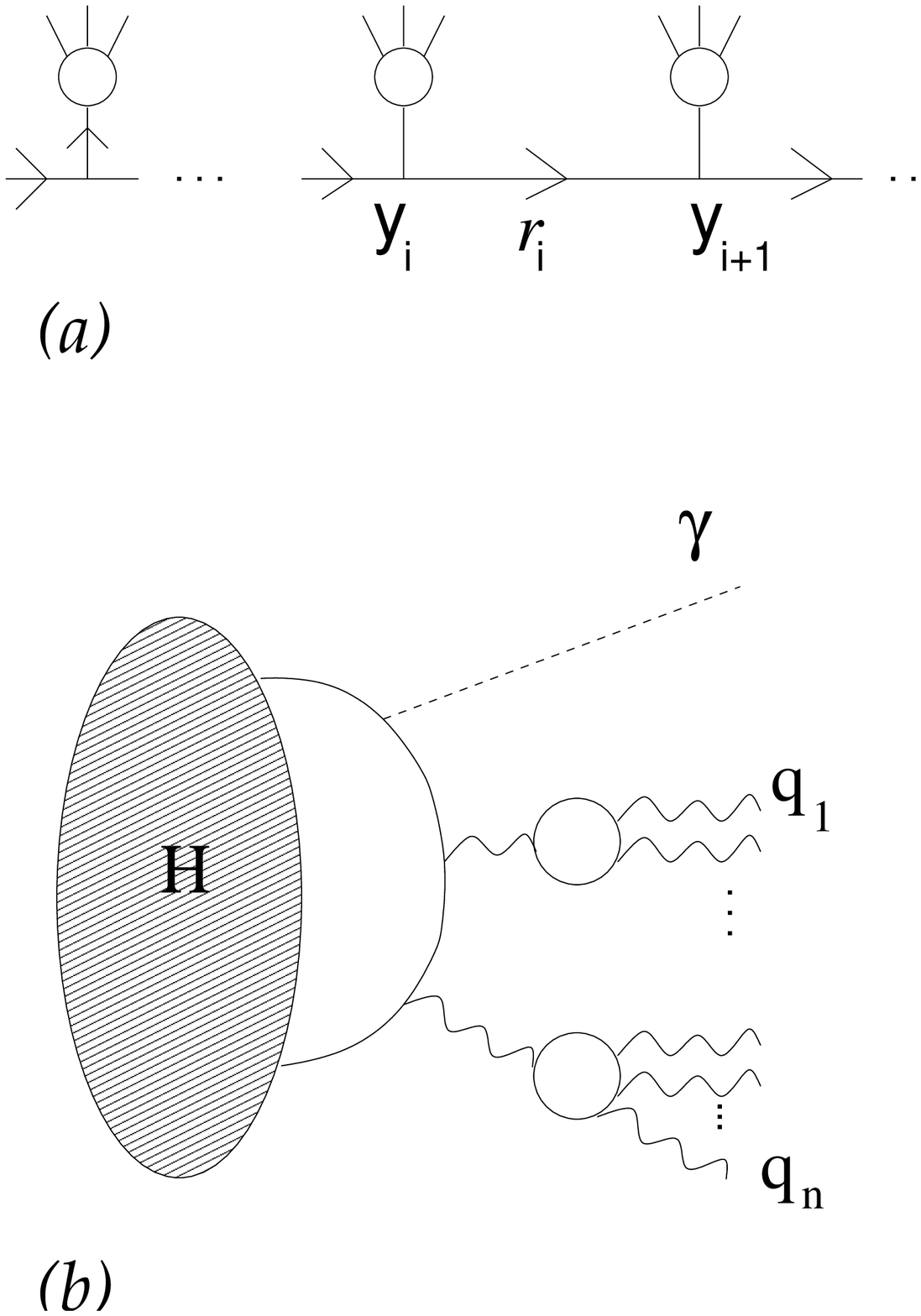 width 8 cm)}
\vspace*{5mm}
\caption{(a) Branching process; (b) final states in 
the quarkonium decay. }
\label{finstate}
\end{figure}

\vskip 0.7 true cm

\noindent  {\it  2.4  Jet mass distributions and inclusive spectrum} 
\vskip .2 true cm

To lowest order, there is only one final state parton in 
the definition (\ref{defstruc}) of  
the jet mass distribution, and  
$J_g$ is just given by a delta function: 
\begin{equation} 
\label{justdelta}
J_g \left( p^2 , k^2 \right) = \delta (k^2) + 
{\cal O} (\alpha_s) 
 \hspace*{0.2 cm} .      
\end{equation}

In general, the  
analysis of the jet branching process allows one to obtain an evolution 
equation~\cite{cttw}  for the jet mass 
distribution. Schematically, this has the form  
\begin{equation}
J (p^2 , k^2) = \delta (k^2) + \int \alpha_s (p^{\prime 2})  
{\cal K} ( 
p^{\prime 2} , k^2) \otimes J  ( p^{\prime 2} , k^2) \;\;\;,    
\end{equation} 
where ${\cal K}$ is a kernel calculable as a power series expansion in 
$\alpha_s$. 
Expressions for the kernel ${\cal K}$ are known to the 
leading~\cite{coh,fad83,dokfad} and 
next-to-leading order~\cite{cttw,jiro}. 
Solving the evolution equation to the leading logarithms 
gives 
\begin{equation}
\label{llaresult}
\int d k^2 \ J_g ( p^2 , k^2) \ \Theta (K^2 - k^2) = 
\exp \left[ \ln \left( {p^2 \over K^2} \right)  f \left( \beta_0  
\alpha_s  \ln \left( {p^2 \over K^2} \right) \right) \right]
 \hspace*{0.2 cm} ,    
\end{equation}
where $\beta_0 = (11 C_A - 2 N_f)/(12 \pi)$ and $f$ is the 
function
\begin{equation}
\label{llafun}
f(x) = - { C_A \over { 2  \pi  \beta_0}} \ {1 \over x} \ 
 \left[ (1-2  x) \ \ln(1 - 2 x) - 2 \ (1-x) \ \ln(1-x) \right] 
 \hspace*{0.2 cm} .     
\end{equation}

It is also useful to introduce 
the double logarithmic approximation. This is  defined by 
 expanding the exponent in 
the right hand side of Eq.~(\ref{llaresult}) 
 to the first order in $\alpha_s$. In this 
 approximation, using $f(x) = - C_A  x / (2  \pi  \beta_0) 
 + {\cal O} (x^2)$,  we get  
\begin{equation}
\label{doublog}
\int d k^2 \ J_g ( p^2 , k^2) \ \Theta (K^2 - k^2) 
\approx 
\exp \left[ - {{\alpha_s 
} \over {2  \pi} } \ C_A \ \ln^2 \left( {p^2 \over K^2} \right) \right] 
 \hspace*{0.2 cm} .   
\end{equation}

In order to express the inclusive spectrum  in terms of 
the jet mass distributions, we now use the  
 definitions (\ref{insert}),(\ref{defstruc}) in Eq.~(\ref{excl}). 
By taking account of the photon phase space factor, the inclusive spectrum 
 takes the form 
\begin{eqnarray}
\label{branchfact}
{1 \over \Gamma} \ 
{{d \Gamma} \over { d z} } 
&=& 
 \int  {{ d^4 k} \over {(2 \pi)^3}} 
{{ d^4 k_1} \over {(2 \pi)^3}} {{ d^4 k_2} \over {(2 \pi)^3}} 
%\nonumber\\
%&\times& 
(2 \pi)^4 \delta^4 (P-k-k_1-k_2) 
\delta \left(z - {{2 P \cdot k} \over M^2} \right)
\delta_+ (k^2) 
\nonumber\\
&\times& 
  {\cal M}_0 (P, k_1 , k_2) \ 
  J_g \left( (k_1+k_2)^2 , k_1^2 \right) \ 
  J_g \left( (k_1+k_2)^2 , k_2^2 \right) 
\hspace*{0.6 cm} .     
\end{eqnarray}
Here  ${\cal M}_0 $ is the  amplitude for the 
  first step in the branching, 
 with final state $ \gamma g g$. 
Owing to the constraint set by coherence, 
we can use the approximation   $(k_1+k_2)^2 \gg k_1^2 , k_2^2$ and 
evaluate  ${\cal M}_0 $  by setting 
$k_1^2 = k_2^2 = 0$. 
The explicit 
 expression is readily computed and reads 
\begin{equation}   
\label{invmatrel}
{\cal M}_0 = { {8 \pi^3 M^2} \over {\pi^2 - 9}} \ 
 \left[ {{ 4 (k_1 \cdot k_2)^2 } \over 
 { ( P \cdot k_1 )^2 \ ( P \cdot k_2 )^2 }} + 
 { 1 \over { ( P \cdot k )^2 }} \ \left( 
 { { (P-k_1)^4 } \over { ( P \cdot k_2 )^2 }} + 
 { { (P-k_2)^4 } \over { ( P \cdot k_1 )^2 }} \right) 
 \right] \hspace*{0.6 cm} .   
\end{equation}

We are now going to evaluate the branching formula 
(\ref{branchfact}) explicitly. To  this end, we need to  
 analyze  the phase space of  this formula  
in detail.  This is the object of  the next section.

\vskip 1.5 true cm 

\noindent  {\large \bf  3. Structure of the phase space for 
small jet masses} 
\vskip .2 true cm

We are 
interested  in  the angular-ordered, coherent  region 
\begin{equation}
\label{logregion}
k_1^2 \hspace*{0.1 cm} , \hspace*{0.1 cm} k_2^2 \ll 
(k_1+k_2)^2 \ll M^2 
 \hspace*{0.6 cm} .   
\end{equation}

Consider the phase space element in Eq.~(\ref{branchfact}): 
\begin{equation}
\label{dphi}
d \Phi =  d^4 k \ 
 d^4 k_1 \ d^4 k_2 \ 
\delta^4 (P-k-k_1-k_2) \ 
\delta \left(z - {{2 P \cdot k} \over M^2} \right) \ 
\delta_+ (k^2) 
 \hspace*{0.6 cm} .   
\end{equation}

It is convenient to  express  
this phase space in terms of the 
energy fractions $x_1$, $x_2$ defined in Eq.~(\ref{kindef}), 
the jet  masses 
$ k_1^2 $, $ k_2^2 $, and 
\begin{equation}
\label{s12def}
s_{12} = (k_1+k_2)^2  
\hspace*{0.6 cm} .   
\end{equation}
 
We can use the momentum conserving $\delta$ in 
Eq.~(\ref{dphi}) to integrate 
over  $k$. 
From the 
positivity of the jet energies and invariant masses 
we have 
\begin{equation}
\label{posit1}
0 \leq x_1 \leq 1 \hspace*{0.2 cm} , \hspace*{0.4 cm} 
0 \leq x_2 \leq 1 \hspace*{0.2 cm} , \hspace*{0.4 cm} 
x_1 + x_2 \geq 1 
 \hspace*{0.6 cm}   ,    
\end{equation}
\begin{equation}
\label{posit2}
0 \leq k_1^2 \leq x_1^2 M^2 / 4 \hspace*{0.2 cm} , \hspace*{0.4 cm} 
0 \leq k_2^2 \leq x_2^2 M^2 / 4
 \hspace*{0.6 cm}   .     
\end{equation}
The relation (\ref{s12def}) between $s_{12}$ and the jet momenta   
implies  
\begin{equation}
\label{s12diseq}
k_1^2+k_2^2 + {1 \over 2} x_1 x_2 M^2 - 2 | {\bf k}_1 | 
| {\bf k}_2 | 
\leq 
s_{12} \leq k_1^2+k_2^2 + {1 \over 2} x_1 x_2 M^2 + 2 | {\bf k}_1 | 
| {\bf k}_2 | 
 \hspace*{0.6 cm}     
\end{equation}
with $ | {\bf k}_1 | = \sqrt{ x_1^2 M^2 / 4 - k_1^2}$,  
$ | {\bf k}_2 | = \sqrt{ x_2^2 M^2 / 4 - k_2^2}$.  

 The  phase space element can then be rewritten as  
\begin{equation}
\label{dphibis}
d \Phi =  {{\pi^2 M^2} \over 4} \ 
 d x_1 \ d x_2 \ d k_1^2 \ d k_2^2 \ d s_{12} \ 
 \delta(z - 2 + x_1 + x_2) \ 
\delta (M^2 (1-z) - s_{12} ) 
 \hspace*{0.6 cm} ,    
\end{equation}
with the constraints (\ref{posit1}),(\ref{posit2}),(\ref{s12diseq}). 

 Eq.~(\ref{s12diseq}) selects in general a subspace 
of a complicated form. But we need to  
 consider it in the 
logarithmic region (\ref{logregion}). 
We may use the $\delta$ functions in 
Eq.~(\ref{dphibis}) to integrate over $s_{12}$ and one of the 
energy fractions, say, $x_2$. Then   Eq.~(\ref{s12diseq}) gives 
\begin{equation}
\label{x1constr}
x_1^- \leq x_1 \leq x_1^+ 
 \hspace*{0.6 cm} ,    
\end{equation}
with 
\begin{eqnarray}
\label{x1pm}
x_1^\pm &=& (1 -z/2) \  ( 
1 - z + k_1^2/M^2 - k_2^2/M^2 )  /  (1-z) 
\nonumber\\
&\pm& (z/2) \ \sqrt{  \left( k_1^2 - k_2^2 \right)^2 / M^4 + (1-z)^2 - 2 \ 
(1-z) \left( k_1^2 + k_2^2 \right)/ M^2 } /  (1-z) 
 \hspace*{0.6 cm} .     
\end{eqnarray}

By approximating the constraints (\ref{x1constr}) 
for $ k_1^2$, $k_2^2 \ll 
M^2 (1 -z)  \ll M^2$, we get 
\begin{equation}
\label{appconstr1}
x_1 \greatersim  {{ k_1^2 } \over  { M^2 (1-z) }}  
 \hspace*{0.6 cm} ,    
\end{equation}
and 
\begin{equation}
\label{appconstr2}
x_1 \lesssim  1 - {{ k_2^2 } \over  { M^2 (1-z) }}
 \hspace*{0.6 cm} ,    
\end{equation}
that is, since $1 - x_1 \simeq x_2$ for $z \to 1$, 
\begin{equation}
\label{appconstr2bis}
x_2 \greatersim  {{ k_2^2 } \over  { M^2 (1-z) }}  
 \hspace*{0.6 cm} .     
\end{equation}

Eqs.~(\ref{appconstr1}) and (\ref{appconstr2bis}) 
tell us that in the coherent region
 the phase space 
available for the evolution of each of the jets 
is bounded by the recoiling jet.    
 For $z \to 1$  this bound  
is tighter than the bound from the fragmentation of 
the jet itself, Eq.~(\ref{posit2}).   In the next section we will 
see that the recoil constraint  conspires with 
the angular-ordered form of the jet mass distribution to destroy any  
 logarithmic hierarchy in the inclusive photon spectrum.

\vskip 1.5 true cm

\noindent  {\large \bf  4. The photon spectrum near the endpoint} 
\vskip .2 true cm

We  now put together the results of Sec.~3 for the 
 phase space with 
the  structure  of the spectrum described in Sec.~2 
based on the coherent branching. 

By using Eqs.~(\ref{invmatrel}),(\ref{dphibis}), 
(\ref{appconstr1}) and (\ref{appconstr2bis}) in   
  Eq.~(\ref{branchfact}), we can  rewrite the photon energy spectrum 
 as follows:  
\begin{eqnarray}
\label{unintG}
{1 \over \Gamma} \ 
{{d \Gamma} \over { d z} } &\simeq& 
{1 \over {\pi^2 -9}} \
%\\
%&\times& 
\int_0^1 d x_1 
  \int_0^1 d x_2   \ \rho (x_1 , x_2, z) \ \delta(z - 2 + x_1 + x_2) 
\  \Theta(x_1+x_2 -1) 
\nonumber\\
&\times&   
 \int_0^\infty d k_1^2   \ J_g \left( M^2 (1-z) , k_1^2 \right) 
 \ \Theta(M^2 x_1 (1-z) -k_1^2) \ \Theta(M^2 x_1^2/4 -k_1^2)
\nonumber\\
&\times&  
 \int_0^\infty d k_2^2   \ J_g \left( M^2 (1-z) , k_2^2 \right) 
 \ \Theta(M^2 x_2 (1-z) -k_2^2) \ \Theta(M^2 x_2^2/4 -k_2^2)
\hspace*{0.2 cm} ,  
%\nonumber  
\end{eqnarray}
with $\rho(x_1 , x_2, z)$  given in Eq.~(\ref{tree}). 
The main structural difference with respect to the case of 
jet event shapes in $e^+ e^-$ annihilation 
is that in Eq.~(\ref{unintG}) 
there is   a  two-dimensional integration over 
a distribution $\rho$ in the energy fractions $x_1$, $x_2$.   
 This  comes from the fact that in quarkonia decays 
the parton branching is  probed by a 
 non-pointlike source. 
Eq.~(\ref{unintG})  allows us to 
discuss  the logarithmic behaviors in the endpoint region.  

Observe first that, by substituting into Eq.~(\ref{unintG}) 
the zeroth-order expression 
(\ref{justdelta}) for 
the jet mass distribution $J_g$, we 
recover the lowest order result (\ref{borng}). So at the lowest order 
of perturbation theory Eq.~(\ref{unintG}) gives the correct answer for 
any $z$. At higher orders of perturbation theory, Eq.~(\ref{unintG}) 
gives an  approximation valid in the region of large $z$ with 
leading-logarithm and next-to-leading-logarithm accuracy, provided 
the jet mass distributions $J_g$ are evaluated with corresponding accuracy. 
 Once expanded to the 
next to lowest order in $\alpha_s$, 
Eq.~(\ref{unintG})
can be matched with the 
NLO perturbation theory result~\cite{krae},  and could thus be used 
to obtain improved predictions, valid over a wider range of $z$. 

Eq.~(\ref{unintG}) enables us to see that,  
although Sudakov logarithms are present in the jet distributions 
associated with the decay,  the corrections in $\ln (1-z)$ to the 
photon energy spectrum cancel order by order in $\alpha_s$. 
The  mechanism for  the cancellation can be seen  in 
its simplest form by using the double-logarithmic 
approximation (\ref{doublog}) for $J_g$. 
By expanding 
the right hand side of 
Eq.~(\ref{doublog}) in powers of $\alpha_s$ 
and substituting this into  Eq.~(\ref{unintG}), we note that 
the higher order corrections to $ d \Gamma / dz$ involve 
integrals of the form 
\begin{equation} 
\label{logx1} 
\int_{1-z}^1 d x_1 \ln^k \left( {{ M^2  (1-z) } \over { M^2 x_1 (1-z) }} 
\right) 
=  \int_{1-z}^1 d x_1 \ln^k (1 / x_1)  
 \hspace*{0.6 cm} .       
\end{equation} 
That is,   logarithmic contributions in $x_1$ arise, 
which are important at the kinematic limit $x_1 \to 0$, 
but these never 
give rise to 
logarithms of $(1-z)$ in the photon  spectrum. 
 Terms in   $\ln(1-z)$  cancel 
 because of coherence,  i.e., as a result 
of the  constraint (\ref{appconstr1})   on the jet mass  
 and of the angular ordering (\ref{angord}) in the  branching.

From Eq.~(\ref{unintG}) we can obtain an explicit 
expression for $ d \Gamma / dz$ by keeping track of all the 
leading logarithms in the jet distributions. This is accomplished 
through Eq.~(\ref{llaresult}). Using this formula in Eq.~(\ref{unintG})   
we get  
\begin{eqnarray}
\label{llagamma}
{1 \over \Gamma} \ 
{{d \Gamma} \over { d z} } &\simeq& 
{1 \over {\pi^2 -9}} \
\int_0^1 d x_1 
  \int_0^1 d x_2   \ \rho (x_1 , x_2, z) \ \delta(z - 2 + x_1 + x_2)
\  \Theta(x_1+x_2 -1) 
\nonumber\\
&\times&   
\exp \left[ \ln (1 / x_1) \ f \left( \beta_0 \alpha_s \ln (1 / x_1) \right) 
+ \ln (1 / x_2) \ f \left( \beta_0 \alpha_s \ln (1 / x_2) \right) \right] 
\hspace*{0.6 cm} ,    
\end{eqnarray}
where  $f(x)$ is given in Eq.~(\ref{llafun}). It is straightforward to 
check from Eq.~(\ref{llagamma}), by 
expanding the integrand in powers of $ \alpha_s$, 
 that no logarithms of $(1-z)$ appear in the perturbative expansion 
 of $ d \Gamma / dz$.

The above results indicate that 
the photon spectrum from  the decay of 
  the color-singlet Fock state in the quarkonium 
 is not Sudakov suppressed  by 
higher orders of perturbation theory 
in the endpoint region.  Higher perturbative corrections 
 give 
rise to a constant shift compared to the  leading order answer  
(\ref{borng}).

 The first  derivative of the spectrum, 
on the other hand,   does get logarithmic corrections in  $(1-z)$. 
This can be realized just from Eq.~(\ref{logx1}) by 
taking the derivative with respect to $z$. A singularity  in  
the derivative of the spectrum at $z = 1$ 
 was indeed noted already in  lowest order  (see 
 comment below Eq.(\ref{tree})).   
The leading logarithmic behavior  
 at  one loop  will be   of the type  
\begin{eqnarray}
\label{drvnll}
{ d \over { d z}}  \left(  {1 \over \Gamma} \ 
{{d \Gamma} \over { d z} } \right)  &=&   
\left[ a_0 \ln \left( 1-z \right) + {\mbox{const}}.   + 
{\cal O} (1-z)^1 \right]  
\\
&+&    \alpha_s  
 \left[  a_1 \ln^3 \left( 1-z \right) +  
 {\mbox{subleading}} \;\;  {\mbox{logs}} \;  + 
{\cal O} (1-z)^1 \right] + {\cal O} (\alpha_s^2) 
\nonumber
\hspace*{0.4 cm} .    
\end{eqnarray} 
It is of much interest   to calculate these corrections. Likely, the 
$z = 1$ divergence that appears in the  derivative  of the spectrum 
at any fixed order of perturbation theory will be smoothed out by 
the all-order summation of the logarithms. Note however that this 
calculation will not necessarily involve only the configurations 
 in which 
the photon and the gluon jets are at large relative angle, and the 
jets evolve according to the angular-ordered branching, but also    
configurations 
 in which color is emitted at angles comparable to that of the photon. 

To conclude this section, we  remark  that 
near the boundary the 
decay becomes sensitive to the hadronization process. For the 
inclusive spectrum, 
these effects 
can be factorized in 
nonperturbative shape functions~\cite{korshape,dikeshif,neushape}. 
A very important question concerns the modeling of these functions. 
The Monte Carlo calculation~\cite{Field} provides one such  model, 
 based on parton showering and the assumption of 
 independent fragmentation. Another model is that of 
 Refs.~\cite{field01,consfie}, in which 
 nonperturbative corrections are parameterized in terms of 
  an effective gluon mass $m_g$~\cite{ppmass}. 
We observe that results on the infrared behavior 
such as those presented in this paper 
can be used to investigate  models of hadronization. 
In particular, if the resummed formulas for the photon spectrum 
discussed above 
 are combined with models for the 
 behavior of the 
 running  coupling at low energy scales~\cite{dikeshif,dmw}, 
 they provide an ansatz for the power corrections in the 
 shape functions~\cite{dok98}.  
These power corrections  are the analogues of the 
contributions in $m_g / M$ of \cite{consfie,ppmass}, 
which are 
found~\cite{field01} to be necessary to describe the data in the 
endpoint region.

\vskip 1.5 true cm

\noindent  {\large \bf  5. Conclusions} 
\vskip .2 true cm

The photon spectrum in quarkonia decays is 
a critical issue in QCD phenomenology. Although, 
as was 
 realized early on~\cite{earliest}, 
 for large enough 
masses the decay is dominated by short distances, 
the observed behavior~\cite{cleo,mark2,earlierups} of 
the spectrum at large $z$ is not reproduced by fixed-order 
perturbation theory. In this respect 
the first implication of the results of 
 this paper is    negative:  perturbative resummations 
 do not fix the problem.  
 We have seen in the previous sections that   resummation 
 does not produce a large-$z$ suppression of the spectrum, but a 
 constant shift.

A second, perhaps more general    
 implication is that color coherence effects  in the quarkonium decay 
are important. As we 
have seen,   they change dramatically the 
large-$z$ behavior of the perturbation series compared to what one would 
conclude from  the simple  infrared power counting~\cite{Photia}, leading 
to 
the cancellation of the Sudakov terms. 

Note that 
the comparison of theory with experiment (see, 
e.g., 
 \cite{montfield,cleo})  has 
so far involved the use of models for the 
 parton cascade and hadronization \cite{Field} 
  that do not include coherence. 
In the case of \cite{Field} the emission of color  associated with the  
evolution of the jets takes place 
within a cone whose typical size is $ \theta \sim 1$. But the 
analysis of this paper indicates that in fact, 
since the coherence scale in the jet mass distributions  is not the 
quarkonium mass $M^2$ but rather $(k_1+k_2)^2$,  
destructive interference occurs outside a cone  
with size $ \theta \sim \sqrt{1 - z}$. 
It will be valuable to make Monte Carlo models for 
quarkonium decays that take this into account.

In the endpoint region 
relativistic corrections may 
become important~\cite{rothwise,grekap,muzi}. 
Detailed estimates of 
color octet contributions have recently appeared~\cite{malpet,wolf}. 
In this paper we have observed  
the absence of Sudakov suppression 
for the Fock state of lowest order in the nonrelativistic 
expansion, i.e., a color singlet quark-antiquark pair. 
The mechanism that produces this result 
is  based on the 
   color correlations 
of two gluon jets, and  therefore 
is   not at work in the case of the color-octet  state.  
This means that the 
power counting in the heavy quark velocity and 
in $\alpha_s$ at fixed order is modified by the 
high order behavior  in 
 very different ways for the  color-singlet and 
color-octet channels. 
The relative size of the 
color octet with respect to the color singlet 
in the endpoint region 
will be smaller than 
indicated by the power counting at fixed order.

Quarkonia decays  are  
used to measure  the QCD 
coupling and 
 contribute significantly to the world 
average value of $\alpha_s$~\cite{bethke2k,hinchmanoh}. In fact 
  Ref.~\cite{hinchmanoh} quotes a 
  very precise determination of $\alpha_s$ using this method.  
As we have seen, however, effects  from the endpoint region 
in radiative decays 
 may be quite dramatic  
(cancellations at 
large $z$, lack of angular ordering in current Monte Carlo models, need for 
an all-order power counting for relativistic corrections), 
and have yet to be fully understood.  
 The present  uncertainty on the extraction of 
$\alpha_s$ from these decays is therefore likely to be larger than 
previously estimated.

There has recently been 
progress in next-to-leading-order calculations for the 
photon spectrum. NLO corrections have been computed 
for color octet contributions~\cite{malpet} and 
for color singlet,  direct contributions~\cite{krae}.  
The  missing piece are the corrections to the fragmentation 
terms \cite{mont}, to be used  with the next-to-leading 
fragmentation functions of the photon~\cite{bourhis}. A possible choice for 
future determinations of $\alpha_s$  will be to  use only data 
sufficiently away from the endpoint for NLO perturbation theory 
to be a valid approximation~\cite{wolf,krae}. 
This raises  
the question, though, 
  of where the safe region begins and how much of the data 
one is left with. 
Alternatively, 
note that  
   resummation formulas 
of the type 
presented in this 
paper can be matched and combined with NLO results. This will give   
 improved predictions, which  
could likely be used in a   wider range of photon energies towards 
the peak   region.  

 Hadronization physics, showing up as 
power-suppressed corrections to the spectrum,  becomes a 
dominant effect  near the endpoint. Power corrections are indeed found to 
be essential to describe the data in this region~\cite{field01,consfie}. 
It becomes important to investigate models for the 
nonperturbative shape functions~\cite{dikeshif,neushape,manshape} 
parameterizing these effects. 
The structure of soft color emission discussed in this paper, once   
combined with  an ansatz for the infrared behavior of the 
coupling~\cite{dok98}, may serve to study these models.

\vskip 1.9 true cm 

\noindent  {\bf  Acknowledgments.} I am grateful to S.~Catani 
for 
collaboration on quarkonium physics and for  discussion. Part of 
this work was done while I was visiting the University of Oregon. 
I  am grateful to  J.~Brau, D.~Soper and 
the University of Oregon Center for High Energy Physics 
for their hospitality and support.  
I thank 
G.~Korchemsky and 
M.~Kr{\"a}mer for useful conversations. This research is funded in 
part by the US Department of Energy under grant No.~DE-FG02-90ER-40577.

\vskip .2 true cm

\vskip 1.5 true cm

\end{document}